\documentstyle[12pt,epsf]{article}
\newcommand{\be}{\begin{equation}}
\newcommand{\ee}{\end{equation}}
\newcommand{\ba}{\begin{eqnarray}}
\newcommand{\ea}{\end{eqnarray}}
\newcommand{\baa}{\begin{eqnarray*}}
\newcommand{\eaa}{\end{eqnarray*}}
\newcommand{\lab}[1]{\label{#1}}

\newcommand{\dis}{\displaystyle}
\newcommand{\non}{\nonumber}

\newcommand{\bhat}{\hat{\beta}}
\newcommand{\pbhat}{P_B}
\begin{document}
\title{Prediction of Peptide Conformation\\ by
Multicanonical Algorithm:\\
A New Approach to the Multiple-Minima
Problem\footnote{Submitted to the Journal of Computational
Chemistry.}\\[4ex]}

\author{Ulrich H.E. Hansmann\\[1.5ex]
{\em Department of Physics and}\\
{\em Supercomputer Computations Research Institute} \\
{\em The Florida State University}\\
{\em Tallahassee, FL 32306} \\[1.5ex]
and\\[1.5ex]
Yuko Okamoto\thanks{On leave of
absence from Department of Physics, Nara Women's
University,Nara 630, Japan.}\\[1.5ex]
{\it Stanford Linear Accelerator Center}\\
{\em Stanford University} \\
{\em Stanford, CA 94309}
}
\maketitle
\newpage

\begin{abstract}
We apply a recently developed  method, multicanonical
algorithm, to the problem of tertiary structure prediction
of peptides and proteins.  As a simple example to test the
effectiveness of the algorithm, Met-enkephalin is studied
and the ergodicity problem, or multiple-minima problem, is
shown to be overcome by this algorithm.  The lowest-energy
conformation obtained agrees with that determined by other
efficient methods such as Monte Carlo simulated annealing.
The superiority of the present method to simulated annealing
lies in the fact that the relationship to the canonical
ensemble remains exactly controlled. Once the
multicanonical parameters are determined, only one
simulation run is necessary to obtain the lowest-energy
conformation and furthermore the results of this one run can
be used to calculate various thermodynamic quantities at any
temperature.  The latter point is demonstrated by the
calculation of the average potential energy and specific
heat as functions of temperature.
\end{abstract}
\newpage

\section{INTRODUCTION}

The prediction of tertiary structures of proteins from their primary
sequences remains one of the long-standing unsolved problems
(for recent reviews, see, for example, Refs.~1--4).
The problem amounts to finding the energy
global minimum out of a huge number of local minima separated by high
tunneling barriers.  Within the presently
available computer resources, the traditional methods such as molecular
dynamics and Monte Carlo simulations at experimentally relevant
temperatures tend to get trapped in local minima, rendering the
simulations strongly dependent on the initial conditions.  One of
promising
methods which alleviate this multiple-minima problem is simulated
annealing.$^5$  The method is based on the \lq\lq crystal
forming" process; during simulation temperature is lowered very slowly
from a sufficiently high temperature to a \lq\lq freezing" temperature.
Simulated annealing was used to refine protein structures from NMR and
X-ray data$^{6-8}$ and
to locate the global minimum-energy
conformations of polypeptides and proteins.$^{9-11}$  The
effectiveness of the method was further tested in many
applications.$^{12-22}$
However, the algorithm is not completely free of faults.
There is no established protocol for annealing and
a certain number (which is not known {\it a priori}) of runs
are necessary to evaluate the
performance.
Moreover, the relationship of the obtained conformations to the
equilibrium canonical ensemble at a fixed temperature remains unclear.

A new powerful method which is referred to as
multicanonical algorithm was recently proposed
by Berg {\it et al.}$^{23,24}$  The idea of this method is based on
performing Monte Carlo simulations in a {\em multicanonical}
ensemble$^{23,25}$ instead of the usual (canonical) Gibbs-ensemble.
The canonical distribution for {\em any} temperature can then be
obtained from {\it one} multicanonical simulation run by the re-weighting
techniques.$^{26}$
In the multicanonical ensemble all energies enter with equal
probability so that a simulation may overcome the barriers between
local minima by connecting back to the high temperature states.
Since the
multicanonical ensemble puts the energy on a one-dimensional random walk,
the global-minimum state can be explored with ease.  The method was
originally developed to overcome the supercritical slowing down of
first-order phase transitions,$^{24,27-29}$ but it has also been
tested for systems with conflicting constraints such as spin glasses
$^{30-32}$ and the three-dimensional random Ising model.$^{33}$
The latter systems suffer
from a similar multiple-minima problem and it was claimed that the
multicanonical algorithm outperforms simulated annealing in these
cases.$^{30}$

In the present work we apply the multicanonical algorithm to the
problem of tertiary structure prediction of peptides and proteins.
Since the purpose of this work is primarily to test the effectiveness
of the algorithm, we have studied one of the simplest peptide,
Met-enkephalin.  This peptide is convenient for our purpose, since the
lowest-energy conformation for the potential energy function ECEPP/2
$^{34-36}$ is known$^{37,38}$ and analyses with Monte Carlo simulated
annealing with ECEPP/2 also exist.$^{18,21}$  We shall show that
by running the multicanonical simulation only once we can not only
reproduce the lowest-energy conformation but also obtain the canonical
distribution at various temperatures.

\section{METHODS}

\subsection{Potential Energy Function}

Met-enkephalin has the amino-acid sequence Tyr-Gly-Gly-Phe-Met.
For our simulations the
backbone was terminated by a neutral NH$_2$-- ~group at the N-terminus
and a neutral~ --COOH group at the C-terminus as in the previous works of
Met-enkephalin.$^{10,18,37,38}$  The potential energy function
that we used is given by the sum of
the electrostatic term, 12-6 Lennard-Jones term, and
hydrogen-bond term for all pairs of atoms in the peptide together with
the torsion term for all torsion angles.  The parameters for the energy
function were adopted from ECEPP/2,$^{34-36}$ and
the computer code
KONF90,$^{15,16}$ which is based on Metropolis algorithm,$^{39}$
was modified to accommodate the multicanonical method.  The peptide-bond
dihedral angles $\omega$ were fixed at the value 180$^\circ$
for simplicity,
which leaves 19 dihedral angles as independent variables.

\subsection{Multicanonical Algorithms}

Since the multicanonical algorithm is already described in detail
elsewhere,$^{23}$ we give only a short overview in this subsection.
In the canonical ensemble, configurations at an inverse temperature
 $\bhat \equiv 1/RT$ are weighted with  the Boltzmann factor
\be
{\cal P}_B (E)\ =\ \exp \left( - \bhat E \right) .
 \ee
The resulting probability
distribution is given by
\be
P_{B}(E)\ ~\propto ~n(E) {\cal P}_{B}(E)~,
\lab{pb}
\ee
where $n(E)$ is the spectral density.
Since $n(E)$ is a rapidly increasing
function and the Boltzmann factor decreases exponentially, $\pbhat (E)$
generally has a bell-like shape.  At a finite temperature the value of
$\pbhat (E)$ for low $E$ is smaller by many orders of magnitudes than
the maximum value of $\pbhat (E)$ (see Fig.~1 below).

In the {\em multicanonical} ensemble,$^{23,25}$ on the other hand,
the probability distribution is defined in such a way that a
configuration with any energy enters with equal probability:

\be
P_{mu} (E) ~\propto ~ n (E) {\cal P}_{mu} (E) = {\rm const}.
\lab{pd}
\ee
It then follows that the multicanonical weight factor should have the form
\be
{\cal P}_{mu} (E) ~\propto ~n^{-1} (E)~. \lab{e3}
\ee
In order to define a explicit form of this weight factor, we
introduce two parameters $\alpha (E)$ and $\beta (E)$ as follows:
$^{23,24}$
\be
{\cal P}_{mu} (E) \equiv e^{-B(E)}
= {\rm exp} \Bigl\{- (\bhat +{\beta}(E))E - {\alpha}(E)\Bigr\}.
\lab{mupa}
\ee
Note that for any fixed $\beta (E)$ and $\alpha (E)$ this leads to the
canonical weight factor with the inverse temperature
$\beta = \bhat + \beta (E)$, therefore
the name ``multicanonical''.
{}From Eqs.\ (4) and (5) we have
\be
 e^{-\beta(E) E - \alpha (E)}~ \propto ~P_B^{-1}~,
\lab{mupa2}
\ee
and this equation is used to determine $\alpha (E)$ and $\beta (E)$
as explained in the next subsection.

The standard Markov process (for instance in a Metropolis update scheme
$^{39}$)
is well-suited to generate configurations which
are in equilibrium with respect to the multicanonical distribution.
Since in the multicanonical ensemble all energies have equal
weight, the energy is enforced onto a one-dimensional random walk
(when simulated with local updates) which insures
that the system can overcome any energy barrier.

Since $P^{-1}_B (E) $ is not {\it a priori} known, one needs
for a numerical simulation  estimators for the
multicanonical parameters $\beta (E)$ and $\alpha (E)$. Once they are
determined, one multicanonical run is in principle enough to find
the global minimum and to calculate all thermodynamic quantities by
re-weighting.$^{26}$

\subsection{Implementation of the Algorithm}

In an actual simulation
the parameters $\alpha (E)$ and $\beta (E)$ can be determined
as follows.  We first run a canonical Monte Carlo
simulation at a sufficiently high temperature $\bhat _0^{-1}$. We
approximate $ P_B (\bhat _0,E)$ at this temperature by a histogram
$\tilde{P}_B (\bhat _0,E_i)$ ($i=1,\cdots, N$) where $N$ is the number of
energy bins.  We then determine the mode, $E_{max}$, of the histogram,
where the histogram has its maximum.
By Eq.\ (6) we have
\be
-\beta (E_i)E_i - \alpha (E_i) ~= ~\ln (\tilde{P}_{B}^{-1}
(\bhat _0,E_i)) + {\rm const.} ~\equiv ~y_i~. \lab{e4}
\ee
The parameters $\alpha (E_i)$ and $\beta (E_i)$ can now be obtained,
for example, by connecting two adjacent points $(E_i,y_i)$ and
$(E_{i+1},y_{i+1})$ by a straight line ($-\beta (E_i)$ being the slope
of the line). We restrict ourselves
to the energy range $E \le E_{max}$,
setting $\beta(E) = 0$ and $\alpha(E) = 0$ outside of this range.
If necessary, this procedure is iterated for a few times
until the obtained distribution $\tilde{P}_{mu}(E_i)$ becomes reasonably
flat in the chosen energy range.  Furthermore,
near the ground-state energy we expect to see
this flat distribution drop to zero abruptly in
a step-function like behavior. This is the criterion for the
optimal choice of $\alpha (E)$
and $\beta (E)$.
After determination of $\alpha (E)$ and $\beta (E)$, we make one
long production run.  Note that the transition probability
$w(E \rightarrow E^{\prime})$ for Metropolis criterion is now given by

\ba
w(E \rightarrow E^{\prime}) &=& 1~, ~~~~~~{\rm if} ~\Delta \equiv
B(E^{\prime})-B(E) \leq 0~, \lab{e5} \\
&=& e^{-\Delta}~, ~~~~~{\rm if} ~\Delta > 0~,  \non
\ea
where $B(E)$ is defined in (5).  From this production run one
can not only locate the global-energy minimum but also obtain the
canonical distribution at any temperature $\bhat^{-1}$ for
all $\bhat \ge \bhat_0$.$^{31}$  The
latter is done by the re-weighting techniques$^{26}$ as follows:
\be
P_B(\bhat,E)=\dis{\frac {\dis{\sum_E} ~e^{B(E)-\bhat E}
{}~P_{mu}(E)} {\dis{\sum_E} ~P_{mu}(E)}}~. \lab{e6} \\
\ee

For our study of Met-enkephalin, we first made a preliminary canonical
simulation at $T=1000$ K with $10^4$ Monte Carlo steps.  We iterated
this process four times to determine optimal $\alpha (E)$ and
$\beta (E)$.  We then made one production run with $10^5$ Monte Carlo
steps recording the time series of the energy and the torsion angles.
The CPU time for the production run was $\approx$ 370 minutes
on an IBM RS/6000 [320H] workstation.

\section{RESULTS}

\subsection{Average Energy and Specific Heat}

We analyze the results of the production run by first calculating the
(canonical) probability distributions, average energy, and specific heat
at various temperatures.

\begin{figure}
\vspace{4in}
\includegraphics{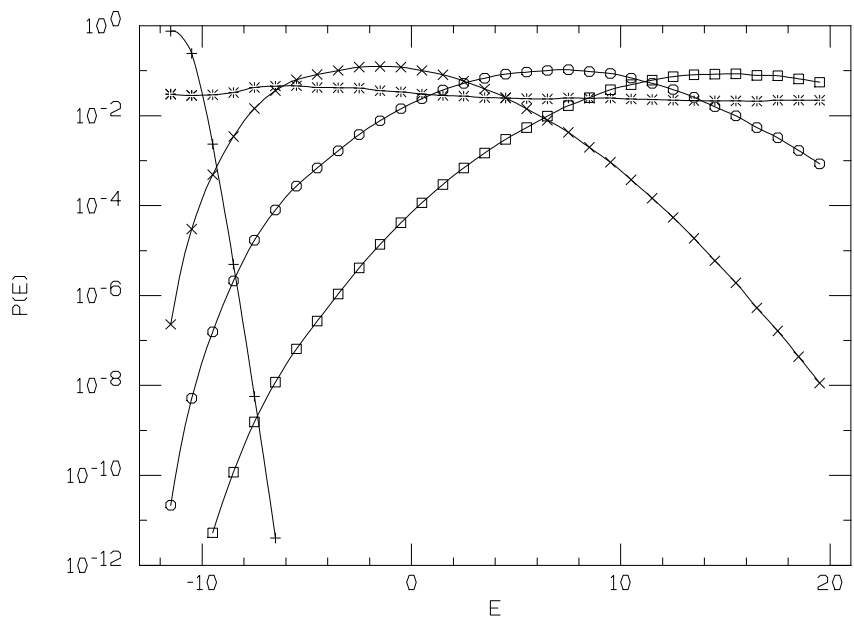}
\caption[fig1]{Probability distributions of
multicanonical ensemble ($*$) and
canonical ensembles at $T=50$ K ($+$), 300 K ($\times$),
500 K ($\circ$), and 1000 K
($\Box$) for Met-enkephalin. }
\end{figure}

In Fig.~1 we show the multicanonical probability distribution
$P_{mu}(E)$ together with the canonical distributions $\pbhat (E)$ at
$T=50$ K, 300 K, 500 K, and 1000 K.  These $\pbhat$ were obtained from
$P_{mu}(E)$ by the reweighting of (9).  Note that $P_{mu}(E)$ is
nearly flat (at least of the same order) throughout the whole energy
range, while $\pbhat (E)$ do vary many orders of magnitude as a function
of energy.  In particular, at higher temperatures ($T=500$ K and 1000 K)
where energy barriers can be easily overcome, it would require canonical
simulations at least $10^{10}$ more simulation time than multicanonical
algorithm to explore the global-minimum energy region with the same
quality of statistics.  This clearly illustrates the advantage of
multicanonical method over the canonical Monte Carlo simulations at a
fixed temperature.

\begin{figure}
\vspace{3in}
\includegraphics{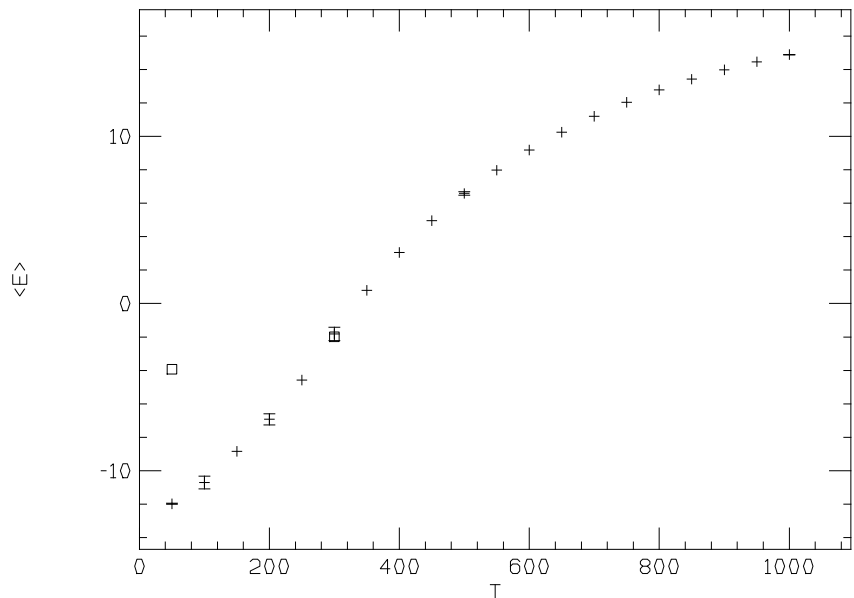}
\caption[fig2]{Average energy of Met-enkephalin
as a function of
temperature
evaluated by multicanonical algorithms.  The results of canonical
simulations at fixed temperatures (50 K and 300 K) are also
plotted ($\Box$).}
\end{figure}

In Fig.~2 we show the average energy as a function of temperature.  This
was again obtained by the re-weighting of (9).  The values vary
smoothly over the whole temperature range. To roughly estimate the errors
of our data, we divided our time series into two bins, the first half
and the second half of $10^5$ Monte Carlo steps.  We calculated
the averages separately for both bins and took their difference as an
estimate for the error, which we included for certain temperatures in
 the figures. The value $\approx -12$
kcal/mol at $T=50$ K is very close to the global-minimum energy obtained
by other methods.$^{18,21,37,38}$
This indicates that the multicanonical
algorithm
avoids being trapped in a local-energy minimum.  In order to illustrate
the effectiveness of the algorithm, we have also listed in the Figure
the values obtained from fixed temperature canonical simulations with
$10^5$ Monte Carlo steps at $T=50$ K and 300 K.  Note that the value
for $T=50$ K is completely off from the multicanonical result, indicating
that this canonical run got trapped in a local minimum.  The value at
$T=300$ K seems in agreement with the multicanonical run.  In fact,
this kind of analysis will tell us how many Monte Carlo steps are
necessary in order that a usual canonical simulation at a certain
temperature may be trusted.

\begin{figure}
\vspace{3in}
\includegraphics{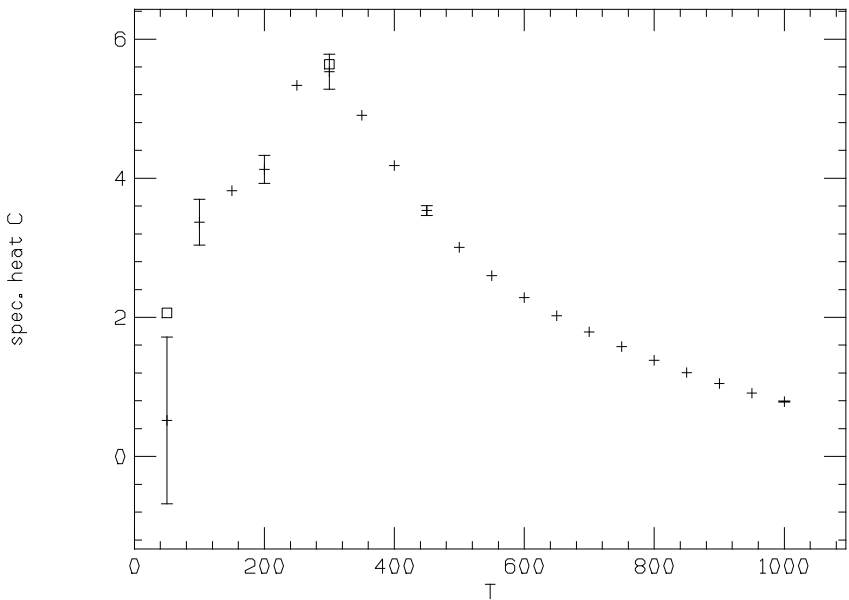}
\caption[fig3]{Specific heat of Met-enkephalin as a function of
temperature
evaluated by multicanonical algorithms.  The results of canonical
simulations at fixed temperatures (50 K and 300 K) are also
plotted ($\Box$).}
\end{figure}

In Fig.~3 we likewise present the \lq\lq specific heat" (per residue),
which is defined by
\be
C = \beta^2 ~\dis{\frac {<E^2>-<E>^2}{5}}~. \lab{e7}
\ee
It has a peak around $T=300$ K, which indicates that this temperature
is important for peptide folding.  The result agrees with the previous
evaluation from canonical simulations at several
temperatures.$^{40}$  The results from the canonical simulations at
$T=50$ K and 300 K also agree roughly with the multicanonical results.
This indicates
that energy fluctuations are not much different whether we do
simulations in the entire conformational space or around a local
minimum.

\subsection{Lowest-Energy Conformation}

During the production run the system reached the global-energy minimum
region in six separate short time spans.  The lowest-energy conformation
within each visit is listed in Table~1 together with the global-minimum
energy conformation (Conformation A in Table~1) obtained by simulated
annealing.$^{21}$  Conformation A has essentially the same structure
as the global-minimum conformation obtained by another method.$^{37}$
The small differences presumably arise
because the peptide-bond dihedral angles
$\omega$ were fixed at the value $180^{\circ}$ in Ref.~21, while
they were allowed to vary in Ref.~37.  Since we use
the same computer code, KONF90,$^{15,16}$ as in Ref.~21 and
fix $\omega$ at $180^{\circ}$ in this work, we compared the present
simulations with the global-minimum conformation of Ref.~21
(Conformation A in Table~1).  We remark that
fixing the $\omega$ angles to the values of
Ref.\ 37 we were able to reproduce essentially the same structure as in
Ref.\ 37.

\begin{table}[tbh]
\begin{center}
\caption[tab1]{Energy and dihedral angles of the lowest-energy
conformations of Met-enkephalin obtained by multicanonical
runs.$^a$}
\vspace{2ex}
\begin{tabular}{|c|r|r|r|r|r|r|r|} \hline
 Conformation  &   A~~ &  1~~  &  2~~  &  3~~  &   4~~ &    5~~ &  6~~\\ \hline
 E [ kcal/mol ]&$-11.9$&$-11.9$&$-12.0$&$-12.0$&$-12.1$&$-12.0$ &$-11.9$\\
\hline
 $\phi_1   $     &  98   & 90    &  91   & 90    & 97    & 96   & 98 \\
 $\psi_1  $      & 154   & 153   & 152   & 154   & 151   & 153  & 156 \\
 $\phi_2 $       &$-161$ &$-160$ &$-157$ &$-161$ &$-158$ &$-161$&$-163$\\
 $\psi_2      $  & 69    & 72    & 64    & 71    & 71    & 68   & 65 \\
 $\phi_3     $   & 65    & 64    & 66    &  63   & 64    & 64   & 66 \\
 $\psi_3    $    &$-93$  &$-95$  &$-92 $ &$-95 $ &$-94$  &$-89$ &$-92$\\
 $\phi_4   $     &$-85$  &$-82$  &$-80 $ &$-77 $ &$-83$  &$-85$ &$-80$\\
 $\psi_4  $      &$-27$  &$-26$  &$-29 $ &$-32 $ &$-30$  &$-31$ &$-29$\\
 $\phi_5 $       &$-83$  &$-81$  &$-82 $ &$-78 $ &$-80$  &$-82$ &$-86$\\
 $\psi_5$        & 142   & 142   & 138   & 137   & 145   & 151  & 147\\
 $\chi^1_1    $  &$-179$ & 179   &$-177$ & 179   & 179   &$-178$&$-176$\\
 $\chi^2_1   $   &$-112$ &$-110$ &$-117$ &$-109$ &$-111$ &$-115$&$-114$\\
 $\chi^3_1  $    & 149   & 144   & 146   & 143   & 149   & 145  & 142\\
 $\chi^1_4  $    & 180   &$-176$ & 178   & 177   & 180   &$-178$& 180\\
 $\chi^2_4  $    &  73   & 79    &  81   & 86    &  79   & 78   & 78\\
 $\chi^1_5  $    &$-65 $ &$-64 $ &$-67 $ &$-67 $ &$-66 $ &$-67$ &$-66$\\
 $\chi^2_5 $     & 180   &$-179$ & 180   & 180   &$-176$ & 180 & 176\\
 $\chi^3_5$      & 179   & 178   & 179   &$-179$ &$-179$ &$-178$&$-178$\\
 $\chi^4_5$      &$-55 $ &$-66$  &$-59 $ &$-62 $ &$-61$  &$-60$&$-57$\\
\hline
\multicolumn{8}{l}{\tiny $^a$ Conformation A is the
lowest-energy conformation obtained by Monte Carlo simulated annealing
(taken from Ref.~21).}
\end{tabular}
\end{center}
\end{table}

In Table~1, Conformations 1--6 are the results at Monte Carlo steps
20128, 39521, 44462, 65412, 89413, and 95143.  Hence, the system reached
the lowest-energy region in every 5000 to 20000 Monte Carlo steps.  The
energies are almost all equal, and the lowest-energy value in the present
work ($-12.1$ kcal/mol) is slightly less than the previous result
($-11.9$ kcal/mol) by simulated annealing.$^{21}$  Most of the
dihedral
angles of the six conformations also agree with the corresponding ones
of Conformation A within $\approx 5^{\circ}$~.
Hence, the conformations in Table~1 are all equivalent.
Note that these six conformations were obtained by only one production
run of multicanonical simulation, while Conformation A was one of 40
Monte Carlo simulated annealing runs (with $10^4$ Monte Carlo steps).
In this respect multicanonical algorithm is superior to simulated
annealing; only one run is required for the former, whereas in the
latter one does not know $a~priori$ how many runs are required and
the convergence must be tested by running at least several times.

\begin{figure}[bth]
\vspace{3.5in}
\includegraphics{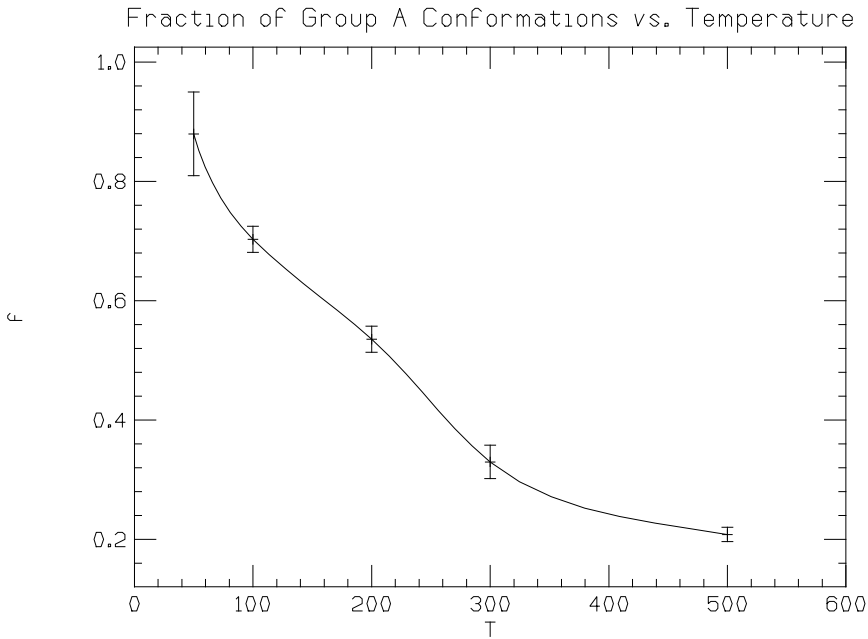}
\caption[fig4]{Fraction of the occurrence of the lowest-energy structure
of Met-enkephalin as a function of temperature.}
\end{figure}
By utilizing the re-weighting of (9), we have calculated the
the fraction in which the lowest-energy conformation exists at various
temperatures (50 K, 300 K, and 500 K).
For this we consider that a conformation is of
the lowest-energy structure if all the 18 dihedral angles
 agree with
those of Conformation A in Table~1 within $\pm 20^{\circ}$.  The
results are shown in Fig.~4.  As expected, at $T=50$ K the peptide
is almost always in \lq\lq ground state".  As the temperature rises,
the conformation is thermally excited and the fraction in Fig.~4
decreases.  However, at $T=300$ K the peptide still stays close to
the \lq\lq ground state" for a substantial amount of time (
$\approx 35$ \%).  This kind of analysis will be useful in understanding
the relation between the conformation with the global-minimum potential
energy and the native conformation around room temperature.

\section{CONCLUSIONS AND DISCUSSION}

In this article we have applied the recently developed multicanonical
algorithm to the problem of peptide conformation prediction.
This method avoids getting trapped in a local minimum of energy function
by connecting
back to high temperature states and enhances in this way the probability
to find the global minimum. This property is exactly what we need for peptide
structure prediction.  We have demonstrated the effectiveness of the
algorithm by reproducing the lowest-energy conformation of
Met-enkephalin.  This was achieved by only one production run of
simulation, whereas another powerful method for overcoming energy
barriers such as simulated annealing usually requires much more runs to
confirm the results.  Furthermore, the multicanonical algorithm can
yield various thermodynamic quantities as a function of temperature
from only one production run.  This was not possible by previous
methods.  To illustrate this property, we have calculated the
average energy and specific heat at various temperatures.

Although our method for the determination of the multicanonical
parameters $\alpha (E)$ and $\beta (E)$ is quite general, it required
about 50 \% of the CPU time spent for the production run.  It is thus
desirable to develop a more efficient method for the determination of
these parameters.  Work in this direction is in progress.

As far as
one is only interested in finding the global minimum, another promising
algorithm would be a related method, {\em random cost optimization}.
$^{41}$
Comparison of the performance of the multicanonical
algorithm
and random cost optimization in the problem of peptide structure
prediction is now under way.

\section*{Acknowledgements}

  One of us (Y.O.) is grateful to the members of Professor Baldwin's
Group, Stanford University School of Medicine, and
Stanford Linear
Accelerator Center for their kind hospitality.
Our simulations were performed on the SCRI cluster of fast RISC workstations.
This work was supported, in part,
by the Department of Energy, by contracts DE-AC03-76SF00515,
DE-FG05-87ER40319, DE-FC05-85ER250000 and by the
Deutsche Forschungsgemeinschaft under contract \hbox{H180411-1}.



\newpage
\begin{thebibliography}{99}

\bibitem{Rev1} H.A. Scheraga,
  {\it J. Protein Chem.}, {\bf 6}, 61~(1987).

\bibitem{Rev2} F.E. Cohen and I.D. Kuntz, in
in {\it Prediction of Protein Structures
    and the Principles of Protein Conformations}, G.D. Fasman, Ed.,
    Plenum Press, New York, 1989, pp. 647--705.

\bibitem{Rev3} M. Karplus and G.A. Petsko, {\it Nature},
  {\bf 347}, 631~(1990).

\bibitem{Rev4} M. Levitt, {\it Curr. Opin. Struct. Biol.},
  {\bf 1}, 224~(1991).

\bibitem{SA} S. Kirkpatrick, C.D. Gelatt, Jr., and M.P. Vecchi,
  {\it Science}, {\bf 220}, 671~(1983).

\bibitem{Nil} M. Nilges, G.M. Clore, and A.M. Gronenborn,
{\it FEBS Lett.}, {\bf 229}, 317~(1988).

\bibitem{Brun} A.T. Br{\" u}nger,
{\it J. Mol. Biol.}, {\bf 203}, 803~(1988).

\bibitem{Brun2} A.T. Br{\" u}nger, M. Karplus, and G.A. Petsko,
  {\it Acta Cryst.}, A{\bf 45}, 50~(1989).

\bibitem{SA1} S.R. Wilson, W. Cui, J.W. Moskowitz, and K.E. Schmidt,
{\it Tetrahedron Lett.}, {\bf 29}, 4373~(1988).

\bibitem{SA2} H. Kawai, T. Kikuchi, and Y. Okamoto, {\it Protein
Eng.}, {\bf 3}, 85~(1989).

\bibitem{SA3} C. Wilson and S. Doniach, {\it Proteins}, {\bf 6},
193~(1989).

\bibitem{SA4} P. Affinger and G. Wipff, {\it J. Comp. Chem.}, {\bf 11},
19~(1990).

\bibitem{RSA0} D.S. Goodsell and A.J. Olson, {\it Proteins}, {\bf 8},
195~(1990).

\bibitem{RSA1} S.R. Wilson and W. Cui, {\it Biopolymers}, {\bf 29},
225~(1990).

\bibitem{KONF} H. Kawai, Y. Okamoto, M. Fukugita, T. Nakazawa, and
T. Kikuchi, {\it Chem. Lett.}, {\bf 1991}, 213.

\bibitem{KONF2} Y. Okamoto, M. Fukugita,
T. Nakazawa, and H. Kawai, {\it Protein Eng.}, {\bf 4}, 639~(1991).

\bibitem{RSA3} M. Fukugita, T. Nakazawa, H. Kawai, and Y. Okamoto,
{\it Chem. Lett.}, {\bf 1991}, 1279.

\bibitem{RSA4} B. von Freyberg and W. Braun, {\it J. Comp. Chem.},
{\bf 12}, 1065~(1991).

\bibitem{RSA5} K.-C. Chou and L. Carlacci, {\it Protein Eng.},
{\bf 4}, 661~(1991)

\bibitem{RSA6} T. Nakazawa, H. Kawai, Y. Okamoto, and M. Fukugita,
{\it Protein Eng.}, {\bf 5}, 495~(1992).

\bibitem{EnkO} Y. Okamoto, T. Kikuchi, and H. Kawai, {\it Chem. Lett.},
{\bf 1992}, 1275.

\bibitem{RSA8} K.-C. Chou, {\it J. Mol. Biol.}, {\bf 223}, 509~(1992).

\bibitem{MU} B.A. Berg and T. Neuhaus, {\it Phys. Lett.}, {\bf B267},
             249~(1991).

\bibitem{MU1} B.A.\ Berg and T.\ Neuhaus,\ {\it Phys. Rev. Lett.},
              {\bf 68}, 9~(1992)

\bibitem{TV}    G.M.\ Torrie and J.P.\ Valleu, {\it J.\ Comp.\ Phys.},
                {\bf 23}, 187~(1977).

\bibitem{FS} A.M.\ Ferrenberg and R.H.\ Swendsen,
              {\it Phys.\ Rev.\ Lett.},
              {\bf  61}, 2635 (1988); {\bf 63 }, 1658(E) (1989), and
              references given in the erratum.

\bibitem{MU2} B.\ Berg, U.\ Hansmann and T.\ Neuhaus,
               SCRI-91-125, to appear in  {\it Phys.\ Rev.\ B}.

\bibitem{MU3} B.\ Berg, U.\ Hansmann and T.\ Neuhaus,
               BI-TP 92/20, to appear in {\it Z.\ Phys.\ B}.

\bibitem{MU4} W.\ Janke,\ B.\ Berg and M.\ Katoot,
               {\it Nucl.\ Phys.}, {\bf B382} 649~(1992).

\bibitem{SG3} B.\ Berg and T.\ Celik,
           to appear in {\it Int.\ J.\ Mod.\ Phys.\ C}.

\bibitem{SG1} B.\ Berg and T.\ Celik, {\it Phys.\ Rev.\ Lett.},
            {\bf 69} 2292~(1992).

\bibitem{SG2} B.\ Berg,\ T.\ Celik and U.\ Hansmann, FSU-SCRI-92-121
              to appear in {\it Europhysics Letters}.

\bibitem{RIM} E.\ Marinari and G.\ Parisi, {\it Europhysics Letters},
            {\bf 19} 451~(1992).

\bibitem{EC1} F.A. Momany, R.F. McGuire, A.W. Burgess, and H.A.
Scheraga, {\it J. Phys. Chem.}, {\bf 79}, 2361~(1975).

\bibitem{EC2} G. N{\'e}methy,
M.S. Pottle, and H.A. Scheraga, {\it J. Phys. Chem.}, {\bf 87},
1883~(1983).

\bibitem{EC3} M.J. Sipple, G. N{\'e}methy, and H.A. Scheraga,
{\it J. Phys. Chem.}, {\bf 88}, 6231~(1984).

\bibitem{Enk} Z. Li and H.A. Scheraga, {\it Proc. Natl. Aca. Sci.,
U.S.A.}, {\bf 84}, 6611~(1987).

\bibitem{Enk2} A. Nayeem, J. Vila, and H.A. Scheraga,
{\it J. Comp. Chem.}, {\bf 12}, 594~(1991).

\bibitem{Metro} N. Metropolis, A.W. Rosenbluth, M.N. Rosenbluth,
A.H. Teller, and E. Teller, {\it J. Chem. Phys.}, {\bf 21}, 1087~(1953).

\bibitem{L91} Y. Okamoto, M. Fukugita, H. Kawai, and T. Nakazawa,
{\it Nucl. Phys. B (Proc. Suppl.)}, {\bf 26}, 659~(1992).

\bibitem{RACO} B.A.\ Berg, \lq\lq Random-Cost-Optimization",
               to appear in {\em Nature}.


\end{thebibliography}
\end{document}